\documentclass[manuscript,screen]{acmart}

\newcommand{\lp}{\texttt{Lake\_Problem\_DPS}}
\newcommand{\wrf}{\texttt{WRF}}
\newcommand{\frb}{\texttt{FRB\_pipeline}}

\title{Reproducible and Portable Workflows for Scientific Computing and HPC in the Cloud}

\newcommand{\cu}{Cornell University }
\newcommand{\cac}{Center for Advanced Computing}

\author{Peter Vaillancourt}
\orcid{0000-0002-1277-7752}
\affiliation{
    \institution{\cu}
    \department{\cac}
    \city{Ithaca}
    \state{NY}
    \country{USA}
}
\email{pzv2@cornell.edu}

\author{Bennett Wineholt}
\affiliation{
    \institution{\cu}
    \department{\cac}
    \city{Ithaca}
    \state{NY}
    \country{USA}
}
\email{bmw39@cornell.edu}

\author{Brandon Barker}
\affiliation{
    \institution{\cu}
    \department{\cac}
    \city{Ithaca}
    \state{NY}
    \country{USA}
}
\email{brandon.barker@cornell.edu}

\author{Plato Deliyannis}
\authornote{REU Student at Cornell University Center for Advanced Computing}
\affiliation{
    \institution{\cu}
    \city{Ithaca}
    \state{NY}
    \country{USA}
}
\email{pad233@cornell.edu}

\author{Jackie Zheng}
\authornotemark[1]
\affiliation{
    \institution{\cu}
    \city{Ithaca}
    \state{NY}
    \country{USA}
}
\email{jz675@cornell.edu}

\author{Akshay Suresh}
\affiliation{
    \institution{\cu}
    \department{Astronomy Department}
    \city{Ithaca}
    \state{NY}
    \country{USA}
}
\email{as3655@cornell.edu}

\author{Adam Brazier}
\affiliation{
    \institution{\cu}
    \department{\cac}
    \city{Ithaca}
    \state{NY}
    \country{USA}
}
\email{brazier@cornell.edu}

\author{Rich Knepper}
\affiliation{
    \institution{\cu}
    \department{\cac}
    \city{Ithaca}
    \state{NY}
    \country{USA}
}
\email{rich.knepper@cornell.edu}

\author{Rich Wolski}
\affiliation{
    \institution{University of California, Santa Barbara}
    \department{Computer Science Department}
    \city{Santa Barbara}
    \state{CA}
    \country{USA}
}
\email{rich@cs.ucsb.edu}

\keywords{Cloud, Scientific Computing, HPC, Automated Deployment, Docker Containers, Terraform, Ansible, Multi-VM MPI}

\ccsdesc[300]{Applied computing~Astronomy}
\ccsdesc[300]{Applied computing~Earth and atmospheric sciences}
\ccsdesc[300]{Applied computing~Environmental sciences}
\ccsdesc[500]{Computing methodologies~Distributed computing methodologies}
\ccsdesc[500]{General and reference~Evaluation}
\ccsdesc[500]{Software and its engineering~Cloud computing}
\ccsdesc[500]{Computer systems organization~Cloud computing}

\copyrightyear{2020}
\acmYear{2020}
\setcopyright{acmcopyright}\acmConference[PEARC '20]{Practice and Experience in Advanced Research Computing}{July 26--30, 2020}{Portland, OR, USA}
\acmBooktitle{Practice and Experience in Advanced Research Computing (PEARC '20), July 26--30, 2020, Portland, OR, USA}
\acmPrice{15.00}
\acmDOI{10.1145/3311790.3396659}
\acmISBN{978-1-4503-6689-2/20/07}

\begin{document}

\begin{abstract}
The increasing availability of cloud computing services for science has changed 
the way scientific code can be developed, deployed, and run.
Many modern scientific workflows are capable of running on cloud computing resources. 
Consequently, there is an increasing interest in the scientific computing community
in methods,  tools, and implementations that enable moving an application to the
cloud and simplifying the process, and decreasing the time to meaningful scientific
results.
In this paper, we have applied the concepts of containerization for portability 
and multi-cloud automated deployment with industry-standard tools to three scientific 
workflows.
We show how our implementations provide reduced complexity to portability of both
the applications themselves, and their deployment across private and public clouds.
Each application has been packaged in a Docker container with its dependencies and
necessary environment setup for production runs.
Terraform and Ansible have been used to automate the provisioning of compute resources
and the deployment of each scientific application in a Multi-VM cluster.
Each application has been deployed on the AWS
and Aristotle Cloud Federation platforms.
Variation in data management constraints, Multi-VM MPI communication, and  
embarrassingly parallel instance deployments were all explored and reported on.
We thus present a sample of scientific workflows that can be simplified using the
tools and our proposed implementation to deploy and run in a variety of cloud environments.

\end{abstract}

\maketitle
\renewcommand{\shortauthors}{Vaillancourt and Wineholt et al.}

\section{Introduction}
\label{sec:intro}
Scientific computing applications often make use of large-scale, high-performance resources
(computers, networking, and storage, such as those provided by XSEDE) to
achieve ``capability'' results -- new scientific results that are made
possible by the capability of the resources.  Because these resources are
expensive to provision and maintain, they are often deployed in a bespoke
configuration that requires highly optimized coding, data management, and
access methodologies to ensure maximal utilization.  

However, there are a number of
scientific workloads that innovate through other forms of computational
scientific exploration and
discovery.  Specifically, researchers investigating new algorithms or
developing new computationally supported processes often require fast
turn-around times (to support rapid prototyping), resource portability (to enable collaboration), 
and maximal developer productivity. 

Cloud computing has evolved as a commercial approach to meeting these goals
for consumer-facing services.  Often, cloud applications have life-cycles
measured in days or weeks that are developed by a geographically distributed
set of collaborating developers whose labor cost has a considerable budgetary
limitation.  Because cloud computing is optimized for web services, however, it has proven
difficult to exploit for scientific workloads (even those driven by developer
productivity and not resource capability).  In particular, the life-cycle of
many scientific codes is long, creating a valuable legacy that cannot easily
be supplanted by new development.

In this paper, we explore the use of cloud computing, Linux
containers~\cite{containers,singularitypaper,docker}, and an automated deployment scheme 
as productivity enhancing technologies for
scientific applications that include or are based on legacy software.
In particular, for many researchers,
the ease of implementing and running software in multiple cloud environments
becomes a key element of leveraging the flexibility and efficiency of the cloud computing paradigm.
As a result, portability and reproducibility of application installation,
deployment, and decommissioning (i.e. the reproducibility of the software
life-cycle) becomes critical~\footnote{Note that this definition of
``reproducibility'' refers to the reproducibility of the software as a
capability available to its user or users and not to numerical reproducibility
across heterogeneous hardware platforms.}.

Containerization software, which provides application software a lightweight virtualized
environment to run in, has recently become a popular strategy for deploying and running
scientific software, portability across different types of systems, and ease of adoption
for researchers\cite{chamberlain2014using}.  Further, software containers coupled with partially or fully 
automated cloud deployment schemes offer intriguing benefits for a wide range of computational tasks
in scientific research, in the form of robust, scalable, and portable
software deployments that can be used during development through production\citep{reviewcloudtools}.

This paper will describe work successfully performed to encapsulate, deploy,
and run three different \textit{existing} scientific workflows --
which are broadly representative of common computational science applications
-- in multiple clouds using automated containerized deployment.
Our system automatically
\begin{itemize}
\item manages the myriad of different possible deployment options available
from computing clouds,
\item configures the cloud-hosted networking to support virtualized parallel application
execution, and 
\item translates the legacy build and deployment mechanisms that accompany
many application (e.g. from a cluster or batch HPC environment) to the
equivalent mechanisms in the cloud.
\end{itemize}
We make use of Docker containerization technology to provide portability and reproducibility, 
and Terraform\cite{terraform} and Ansible\cite{ansible} to deploy, manage, and
provision cloud resources automatically.
In the following sections we describe each of the scientific workflows in detail, their data and computational requirements,
the particular technical details for containerized implementation, how the choice of deployment context affected the implementation, an evaluation of 
software runs performed, and discuss the practical outcomes of the experience, including the benefits and disadvantages of this approach.
Each of these workflows was run on Amazon Web Services (AWS)\cite{aws}, and Aristotle Cloud Federation\cite{10.1145/3355738.3355755}.

The Aristotle Cloud Federation is an NSF-funded project between the Cornell University Center for Advanced 
Computing, University at Buffalo Center for Computational Research, and the University of California
Santa Barbara Department of Computer Science, with the goal of joining cloud computing resources at each
of these institutions in order to develop a federated model for science users to easily access data,
scale research problems using cloud computing techniques, and lessen the time to science of research
teams.  The federated model allows resources to be shared between the Aristotle member clouds, including
individual data sets, access to specialized software, and access to site-specific resources.  By leveraging
the strengths of each of the member institutions, the overall cloud is able to provide larger overall
scale and more resources than each of those institutions separately.  Use cases described below
are largely the result of collaborations between the Aristotle Science Team members and the Infrastructure
group which drove the requirements for containerized applications.

\section{Scientific Workflows}
\label{sec:workflows}
The following scientific workflows (selected from Aristotle Cloud Federation
Science Use Cases~\cite{Aristotle-Use-Cases}) represent
a broad range of scientific disciplines.  Each case represents a user community
that seeks the potential productivity gains offered by cloud computing.
At the same time, these three examples cover some of the
common challenges encountered when moving scientific code to the cloud.  In section \ref{sec:lake}, we discuss a
message passing interface (MPI) application,
called \lp, used in environmental science research that typically utilizes multiple nodes with low amounts of MPI communication.  
Section \ref{sec:wrf} is a commonly used application in HPC for atmospheric sciences called \wrf, which utilizes higher levels of
MPI communication.  Our final workflow in \ref{sec:frb} does not require MPI or even communication between nodes, but instead
requires high data throughput for processing large radio astronomy datasets.

\subsection{\lp}
\label{sec:lake}
In environmental science, complex systems are studied computationally using the Many-Objective Robust Decision Making (MORDM) framework, 
which enables understanding when decisions must be made while these systems are changing\cite{KASPRZYK201355}.
There is a classic problem -- called the shallow lake problem -- where a town with a lake must make policy decisions
about pollution that will impact the lake's water quality as well as the town's economy\cite{classic-lake}.  Julianne Quinn et. al.
demonstrated the \lp~software\cite{original-repo}, based on the MORDM framework, in solving this problem
using Direct Policy Search (DPS)\cite{DPS} and intertemporal open loop control\cite{QUINN2017125}.  

The software was originally run on an HPC cluster, and utilizes low
amounts of MPI communication throughout the run.  There are no external input data requirements to verify functionality,
so the Aristotle Cloud Federation Science Team was able to begin with a fork of the \lp~software repository to containerize, deploy,
run, and evaluate the software in the cloud environment\cite{lp-repo}.  To
effect an automated cloud deployment, our team translated the legacy cluster submission
scripts from PBS to 
Python, and added the environment initialization to the containerization step.  In order to reproduce the results of 
Quinn et. al., we ran the DPS and intertemporal optimization routines,
performed a re-evaluation, and then generated the figures for comparison to
those generated by an unmodified run.

\subsection{\wrf}
\label{sec:wrf}
The weather forecasting community has historically valued large data sets for predictive power, and utilized analyses of past similar situations to both infill data
and make future projections. The computational Weather Research and Forecasting (WRF) Model\cite{WRF} is popular for weather simulation with a long history 
of development and use by the National Center for Atmospheric Research (NCAR), contributors, and consumers.  The software is widely used by a community of 
more than 48,000 researchers across 160 countries to produce a wide variety of results ranging from contributions to real-time weather prediction, long term climate 
simulations, large-scale low resolution idealized physics simulations, and small-scale high resolution detailed physics simulations leveraging large quantities of 
observational grounding data as model inputs.

\subsubsection{Workflow Complexities}
Numerical computation, data input,
and data output
can all grow large very quickly when simulating detailed physics at high grid resolutions or long timescales.  The communication of intermediate results at 
grid boundaries -- necessary to advance simulation steps at sufficient accuracy -- can also place a burden on network capacity. Therefore, to achieve desired 
modeling fidelity, WRF is thus commonly run on resources with an abundance of computational capacity, disk storage, and network throughput.
Common technologies used to meet these needs include managed compute cluster resources with provided Fortran compilation guides and packages to 
facilitate efficient numerical simulation. Network communication is facilitated by MPI libraries which may be optimized for 
low latency use of specialized network hardware. Disk storage may be fulfilled by high capacity Lustre distributed file system hosting.

The specific WRF model we choose requires parallel execution across compute resources to allow for faster and more detailed numerical 
grid simulation of weather properties, namely of interest to simulating wind speed near wind turbine farms at high spatial and temporal resolution.
Useful simulation data for climate observations include wind speed and temperature, as well as dependent measures such as estimated wind turbine 
power production.  In order to obtain these measures in a reasonable timeframe, it is necessary to leverage large computational resources to quickly 
and accurately simulate many numerical values over grids of varying density, with associated network communication at tile boundaries and large demand 
for disk storage both for tile boundary grounding conditions derived from data input as well as intermediate result storage.  Consequently, this scientific workflow
represents an example of resource-intensive HPC applications and the
challenges they present to effective cloud deployment.

NCAR provides many public data sets, analysis tools, and regression suites
suitable for confirming the validity of the numerical simulations 
produced by the model.  We use these regression tests to validate the
correctness of our cloud WRF executions.

\subsubsection{WRF Test Run Characteristics}
A sample WRF version 4.0 run using an NCAR regression test Docker build was run using 1.3GB
of novel weather Global Forecast System data published by NCEP. Geographic
reference data for grid domain preprocessing and high resolution physics totaled
30GB. This sample was run inside a Docker container on a 4 virtual CPU AWS
instance. The observed runtime was 9 minutes 20 seconds and should be scalable
to moderately larger data sets of similar nature. Similar past simulation runs
executing
private builds and data have leveraged Docker to execute for long time periods
on Aristotle Cloud Federation and XSEDE Jetstream cloud resources.  

\subsection{\frb}
\label{sec:frb}
Fast Radio Bursts (FRBs) are astrophysical phenomena that occur as transient high energy pulses or bursts in radio astronomy data.  
FRBs are expected to occur thousands of times per day, but confirmed detections of unique sources are below a hundred\cite{Cordes_2019} 
since the first recorded detection occurred in 2007\cite{Lorimer_2007}.  Since radio telescopes are on the earth's surface, radio astronomy
data is plagued by large amounts of Radio Frequency Interference (RFI), which can block or distort signals, making transient signals like FRBs
even harder to detect, despite large quantities of data available to search.

\subsubsection{Detecting FRBs}
A standard data presentation in time-domain radio observations is the dynamic spectrum, a plot of intensity vs. time and frequency. Typically, the time of arrival of an 
astrophysical radio transient is longer at lower frequencies due to dispersion by plasma in the interstellar medium. In particular, the time of arrival is 
proportional to the inverse square of frequency and the dispersion measure (DM), a constant equal to the integral of electron density in the interstellar 
medium along the line of sight. Thus, in a dynamic spectrum, radio transients exhibit a characteristic quadratic shape. A number of existing transient 
search techniques are based on de-dispersing dynamic spectra using a bank of several plausible DMs, flattening into a time series, and performing 
matched filtering with the time series. These methods are well-tested and have been reliable in discovering new FRBs in the past decade. 

Exploring new detection techniques that have the potential to offer advantages in accuracy or computation cost is also of great interest to astronomers. 
In the past three years, researchers have conducted successful transient searches by applying multi-layer convolutional neural networks to dynamic 
spectra \cite{Connor_2018}\cite{agarwal2019deeper}. The "Friends-Of-Friends" (FOF) algorithm is a straightforward and efficient way to locate 
radio transient candidates by identifying and characterizing clusters of high signal pixels in the dynamic spectrum directly.

\subsubsection{Software}
For this scientific workflow, the software is called the \frb\cite{FRB_Pipeline} and was developed by the Aristotle Cloud Federation Science Team in collaboration 
with Cornell researchers established within the radio astronomy community.  Thus, the software was designed with the computational needs
of the science in mind, as well as the flexibility to expand as new methods of radio transient detection develop.
The \frb is a customizable scientific software package written in Python 3 designed to simplify the process of combing large datasets -- from any of a variety 
of radio telescopes -- to detect FRBs.  The package enables flexible use of established methods to filter RFI, detect candidates, and determine viability of 
candidates as well as the availability of new methods, or even addition of customized methods by the user.  A commonly used package within radio astronomy,
called PRESTO\cite{PRESTO}, is a dependency, and newer methods such as our FOF algorithm are included as well.

\subsubsection{Outline of the FOF Algorithm}
\begin{enumerate}
  \item Average the raw dynamic spectrum
  \item Compute the root mean square (RMS) background white noise, using an iterative method that discards outlier pixels until a convergence threshold is reached
  \item Mark each pixel with signal greater than a constant parameter $m_1$ times the RMS background noise
  \item Group the high-signal pixels marked in (2) in close proximity -- defined as being within a constant number of time bins and constant number of frequency bins (parameters) -- together to form clusters, keeping those with a total intensity higher a given threshold  $m2$
  \item For each cluster, compute and record the following metrics:
      \begin{itemize}
          \item $N$ - number of pixels
          \item Cluster signal-to-noise ratio (SNR) - mean pixel SNR $\times~N$
          \item Signal Mean/Max - mean/maximum pixel intensity
          \item Pixel SNR Mean/Max - signal mean/maximum divided by RMS background noise
          \item Time Start/End Bin - beginning/end in time domain
          \item Frequency Start/End Bin - beginning/end in freq. domain
          \item Slope - orthogonal distance regression linear best fit slope
          \item DM - physical dispersion measure, from quadratic best fit with orthogonal distance regression
      \end{itemize}
  \item Using either the linear fit or the quadratic fit, group the clusters, and extrapolate each cluster across the entire dynamic spectrum to form "superclusters"
  \item Output: a text file containing a list of candidates (clusters) and their metrics that can be sorted by any statistic, and a plot of each section of dynamic spectrum with clusters highlighted
  \item Plot the top candidates
\end{enumerate}

\subsubsection{Space \& Time Requirements}
Smoothing and averaging reduce the variation in intensity between adjacent pixels, which is essential for FOF to work properly. Additionally, averaging reduces 
the size of the data by a significant factor (typically order 100), vastly reducing the computation time of FOF and other search algorithms. However, smoothing 
does take significant computational time. The computational complexity of pure averaging is $m\times n\times k$, with $n,m$ the number of time and frequency 
bins respectively in the dynamic spectrum, and $k$ the number of pixels averaged together. For example, if smoothing with a 2-D Gaussian filter followed by 
decimation is used, a computational complexity of $nm\times\log(n)\log(m)$ is achieved using fast FFT based convolution. While smoothing has a significant 
cost, savings made running FOF on data whose size is multiple orders smaller than the raw data are essential. The FOF algorithm itself has a computational 
complexity of $m\times n$, with $n$ the number of time bins and $m$ the number of frequency bins in the dynamic spectrum. Both computing the RMS 
background noise and comparing each pixel's signal to the first threshold for the bulk of computation time, while the two least squares regressions computed 
for each cluster are relatively insignificant.

\subsubsection{Advantages \& Disadvantages}
In contrast to techniques using de-dispersion and matched filtering, FOF is completely agnostic to signal shape. On one hand, this means that FOF will have 
no trouble identifying astrophysical signals with \textit{any} DMs, but also that FOF is vulnerable to a high rate of false positives due to RFI. Additionally, because
the DM of a given signal is unknown a priori, de-dispersion and matched filtering must be performed on a large set of trial values, which entails a computational 
complexity equal to the number of trial DMs times the $n\log(n)$ time of matched filtering. In many cases, FOF will be faster than these methods. For example, 
for the analysis in which FRB121101, 5016 trial DMs between 0 and $2038\text{pc}\:\text{cm}^{-3}$ (twice the expected maximum galactic DM) were used by 
Spitler et al. \cite{Spitler_2014}, while FOF is effective on the same dataset averaged to only 100 frequency bins.

\subsubsection{Data}
The Breakthrough Listen (BL) project is a comprehensive search for extraterrestrial intelligence using radio and optical telescopes. The BL target list includes 
nearby stars and galaxies, as well as other peculiar astrophysical sources broadly termed “exotica”\cite{2019PASP..131l4505L}.  As part of the latter category, 
BL observed the first discovered repeating fast radio burst, FRB 121102, for 5 hours at 4 - 8 GHz using the Robert C. Byrd Green Bank Telescope. Using the 
GPU-optimized software \texttt{HEIMDALL}\cite{HEIMDALL} to perform dedispersion and matched filtering, Gajjar et al.\cite{2018ApJ...863....2G} detected 
21 FRBs within the first hour of observation. Zhang et al.\cite{2018ApJ...866..149Z} subsequently applied supervised machine learning on the same data to 
identify 93 pulses with <2\% false positive rate. The large number of FRB 121102 pulses detected in this BL observation, together with the completeness of 
FRB detections by Zhang et al. render these data as ideal testbeds for evaluating the performance of our FOF algorithm.  

Furthermore, the dataset containing the detected FRBs is publicly available\cite{frb-data1}\cite{frb-data2}, with sufficiently large file sizes to demonstrate 
the power and flexibility of the cloud to scale deployments to meet data processing needs.
Since there are large amounts of data to process, with a variety of possible processing methods, this scientific workflow lends itself well to an 
embarrassingly parallel implementation.  When deployed to a cluster, the software does not require communication such as MPI, but it does 
require careful data management.

\section{Technical Implementation}
\label{sec:technical}

Due to the aforementioned differences in the scientific workflows, not all of the details of implementation are the same.  
However, the core steps of the process were the same, and we detail where they differed in the coming sections.  
In general, for each workflow:
\begin{itemize}
    \item A Docker container was built to enable portability and reproducibility
    \item Data was stored in locations easily accessible to cloud computing virtual machines (VMs)
    \item Compute VMs and communication networks were deployed with Terraform
    \item Secure shell (ssh) keys for communication between VMs were configured with Ansible
    \item Application containers and associated setup were deployed to multiple cloud VMs using Ansible
    \item Output data was staged on remote storage for user retrieval
    \item All compute resources were decommissioned using Terraform to curtail ongoing cost
\end{itemize}

All of these applications are non-interactive and batch-style, utilizing a single container image for each node (or VM).  Therefore,
they do not require much or any orchestration of interacting services. 
The utilized data storage was commonly simple object storage such as Amazon S3\cite{aws-storage}
or a deployed NFS server on an additional cloud VM when a file system mount was needed by the application.
These compute runs were small-scale (relative to large jobs on an HPC system) and intended as 
proof-of-concept for the applications.  Therefore, no more than 3 maximum size VMs
were provisioned per cluster in our testing runs performed for this work. Other deployments have been made up 
to 8 or more VMs, but a thorough study of maximum effective cluster size or the point of decreasing gains
at scale due to communication overhead is deferred to future work. Reproducibility was gauged by application 
completion on basic testing scenarios, and was not thoroughly evaluated with an eye to numerical precision 
errors nor subtle kernel differences that Docker cannot eliminate, being bound to run on the VM kernel as deployed.

\subsection{Containerization}
\label{sec:containers}

Containers are OS-level virtualization: a single OS kernel can
host multiple isolated environments. Containers are lighter weight than using
hypervisor-based VMs wherein each VM runs a separate kernel,
but comes with the obvious restriction that all containers on the host
must use the same OS kernel and the same version of the
kernel. But in many cases, this has been a more than acceptable
trade-off. While container technology originated
from Solaris Zones\cite{10.5555/1052676.1052707} and
FreeBSD Jails\cite{Kamp00jails:confining},
the prevalence of Linux on commodity cloud hardware has
synergistically aided in creating a convergence of Linux
containerization technology.

Docker has not only been
responsible for both the popularity of Linux containers in the
industry, it has seemingly expanded the definition
of a container. The colloquial definition now includes the ability
to distribute and deploy applications
with minimal configuration: i.e., everything is self-contained within
the container. Singularity,
another Linux container technology, targeted HPC users by trading
security concerns: Singularity requires the user to grant an application
container access to all of the user's files rather than running a container
virtualization service as a privileged user, as is the case with Docker. In so doing,
many of the traditional notions of a container were further broken
down, though Singularity has optional parameters
to enable isolation
\cite{RunningServicesSingularitycontainer30documentation-2020-01-13}.

Other container technologies exist for Linux as well, including that which is
provided by a core service of most Linux distributions,
systemd\cite{systemdnspawn-2020-02-05}.
However, \verb!systemd-nspawn! containers typically do not include the more
modern connotation of a container being a packaged application. But,
it has been used by other technologies for this purpose, such as
nix-containers \cite{NixOSmanual-2020-02-06_containers}, a
container technology for the
NixOS \cite{nix,nixthesis,nixhpc} Linux
distribution that allows packages to be shared from the host's
package-store. While we have not yet used Nix containers in this work, we have
containerized Nix within Docker, which affords its own advantages.

Nix provides a high degree of reproducibility due to package definitions being
carefully check-summed for any sources of differences, e.g.:
URL change of binary or source blobs used for the package, checksum differences in
the binary or source blobs, version changes, configuration changes,
semantic changes in the package definition (i.e. Nix expression) --- such as build
or runtime configuration, or any such changes in dependencies of the packages.
Once a nix expression is written, it can then be shared for use within other Nix projects,
without the need to worry about how to integrate it into a container definition file.
Additionally, by using Nix, the environment could easily be run bare-metal on
NixOS or as a Nix container in the future.

\subsubsection{Workflow Specifics}
For this work, specialized Docker containers were developed for the \lp~and \frb~applications, but
NCAR WRF has a publicly available Docker Container for WRF including a regression test\cite{WRF-Docker}.  
The \texttt{Lake\_Problem\_DPS} container makes full use of Nix within Docker to ensure reproducibility,
even using a Nix expression to simplify the process of including proprietary software within
the container without publicly sharing the software in a public GitHub repository\cite{lp-repo}.  The \frb~
container was based upon a container developed for North American Nanohertz Observatory for 
Gravitational Waves (NANOGrav)\cite{NANOGrav}\cite{nanopulsar}, but updated for our work\cite{our-nanopulsar}\cite{mi-container}.

\subsection{Automation}
\label{sec:auto}

Terraform is an open source tool used for infrastructure management and provisioning developed by HashiCorp. 
The common use case for Terraform is managing resources on multiple cloud infrastructure providers -- such 
as AWS and Google Cloud -- with minimal differences in scripts.
Terraform is under active development and supports a variety of providers including AWS, 
Google Cloud Platform (GCP)\cite{gcp}, Microsoft Azure\cite{azure} , 
and OpenStack infrastructure providers which includes XSEDE Jetstream, 
Aristotle Cloud Federation, and the Cornell CAC Red Cloud on-premise hosting environments.
Terraform uses the HashiCorp Configuration Language (HCL)\cite{HCL} to automate the deployment of various cloud resources among
different cloud vendors. Terraform does this in a declarative manner, meaning that cloud
resource states are written in a Terraform file and Terraform attempts to create the declared resource or
modify the resource into the declared state. It manages the existing resource using metadata created from running 
a Terraform configuration file. 

Ansible is an open source tool used for software provisioning, configuration management, and automation. Ansible 
uses YAML to write configurations. Similar to Terraform, Ansible is declarative, though it can also perform 
operations procedurally. An Ansible YAML file declares states of various Ansible modules which are subsequently 
set up on the remote machine. Compared to using scripts for configuration management, Ansible can achieve the
same things scripts can on multiple machines or VMs in parallel, which is advantageous to cluster management. 

We used Terraform for provisioning and infrastructure management, and Ansible for configuration management. 
Concretely, Terraform was used to create the VMs and networks while Ansible was used to set up Docker
application containers including the associated environments for the science workflows to run on VMs and
to issue commands to initiate and control science runs. To create a cluster 
with communication, Terraform first sets up a single base VM with a custom network configuration. From the network 
configurations, the VM can only receive ssh traffic from a specified IP and TCP connection with another VM on the 
same network. There is no restriction on how the VM can send traffic. Next, Ansible imports the Docker containers. 
Afterwards, Terraform creates copies of the VM to form a cluster. Finally, Ansible sets up the VMs in parallel for OpenMPI communication.

It is important to note that while the Ansible script is portable across different cloud infrastructure providers,
the Terraform script is not. Ansible requires the IP addresses of the VMs while Terraform resources are dependent 
on the cloud infrastructure providers. Generally, to use Terraform on the various providers requires some form of
credentials and slight modifications to the Terraform script.  By using simple VM hosting with standard OS images 
rather than provider-specific services, our infrastructure level scripting is easily portable even though it requires 
some provisioning code translation to adapt to different underlying cloud providers based on commonly available 
templates, including our own new public examples.

\subsection{Multi-VM MPI}
\label{sec:mpi}

After provisioning server and network resources, we configure them using both Ansible scripting and Docker container 
deployment. A simple way to think of the process is that Terraform creates a server we can access, Ansible installs libraries 
we need (much like user shell commands issued over ssh), and Docker will fetch a particular application bundle to be run.

For on-demand MPI clusters in particular, Ansible scripts perform some additional cluster level configuration, such as pushing 
a list of cluster hosts to each member upon setup. The Docker images we created to run WRF, for example, on the on-demand 
MPI cluster similarly have cluster level host configuration injected by a scripted build process to lower the burden of manual user 
setup on each cluster deployment. Ansible can then perform scripted cluster tests to verify success of all deployed components 
and successful networked science code execution. Notably, the entire process can be written as code or templates and run from 
Terraform commands on one researcher workstation. These concepts are similar to functions provided by another popular technology 
for cluster deployment -- Kubernetes -- which specializes in deploying redundant web applications across multiple hosts and 
networks to provide high availability and uptime for diverse workloads. However, due to MPI communication requiring long-lived 
guaranteed hosting and our current focus on a small set of related applications per cluster deployment, neither the additional 
user complexity of scripted Kubernetes setup nor the monetary and researcher familiarity costs associated with hosted Kubernetes 
solutions sacrificing portability are justified under the stated goals of this work.  By using simple and standard technologies summoned 
by a small set of scripts, users can easily create on-demand MPI clusters to achieve science results.

Default usage of Terraform and Ansible examples will create network resources and servers and then configure them to perform 
useful work, but in some cases advanced recurring communication between server instances provides for a new classes of applications 
to be hosted for efficient computation. Multiple VMs can be organized into clusters running message passing interface (MPI) applications 
to deliver high-performance computations commonly associated with large managed hardware clusters. Although we use cloud providers 
with commodity network and storage services as well as hardware level hyperthreading or time sharing in this work, options exist to pay 
a premium for specialized hardware appliances, dedicated hardware, and bare metal native execution with varying amounts of increased 
configuration complexity. Here we defer detailed cost, complexity, and performance analyses to future work and recount our experiences 
developing and deploying on-demand OpenMPI clusters with real scientific research applications on basic cloud provider offerings. As noted 
above, the technologies used  -- namely Terraform, Ansible, and Docker -- are open source or publicly available, and were chosen to 
facilitate smooth researcher user experience, which can be further facilitated by referencing our own published examples.

Our primary on-demand MPI application target for cluster deployment is the NCAR WRF Model widely used for weather simulation. 
This application has intense compute, networking, and storage demands that lend well to scaling different grid tiles of simulation
time steps to multiple VMs with periodic communication of intermediate results and grounding grid tile boundary conditions against 
provided observational data.

\subsubsection{User Experience}
\begin{enumerate}
  \item User checks out deployment code and configures their desired cloud provider credentials and tools, and sets cluster size
  \item User installs Terraform and Ansible with provided commands to perform deployment
  \item User executes deployment which creates a cluster of the desired size on the cloud provider and runs short tests
  \item User may use deployment script outputs to access the cluster for specific manual application runs
  \item User copies out result data and cleans up cluster using Terraform destroy command
\end{enumerate}

Cluster deployment including resource provisioning and server instance configuration is entirely automated, triggered by the
user calling a single Terraform command, reviewing the proposed changes, and choosing to execute. 

\subsubsection{Network and Virtual Machine Image}
In more detail, the cluster provisioning automates the following processes:
\begin{enumerate}
  \item Terraform reads the resources desired from the appropriate cloud provider template folder 
  \item Terraform reads the provided cloud provider credentials
  \item Terraform plans the resources it must create, namely networks, security groups, and VM instances
  \item Terraform requests approval to create the resources, which will incur potential costs
  \item Terraform uses an underlying provisioning API to create resources, starting with networks
  \item Terraform creates a VM where cluster software packages will be installed
  \item Ansible waits for the VM to come up, then installs software packages needed by all cluster nodes, 
           including Docker images with scientific application code
  \item Terraform tears down the VM to ensure a clean disk snapshot
  \item Terraform is notified of Ansible completion and takes a server image
\end{enumerate}

\subsubsection{Cluster Virtual Machines and Network}
\begin{enumerate}
  \item Terraform creates more VM instances as above using the server image as a base copy
  \item Ansible gets IPs of created cluster VMs
  \item Ansible builds cluster specific Docker images with cluster info, and builds in ssh configuration for later use with MPI
  \item Terraform creates NFS server for the cluster (if needed by the application)
  \item Ansible confirms NFS mounts on each host (if needed by the application)
\end{enumerate}

\subsubsection{Application Runs and Cleanup}
\begin{enumerate}
  \item Ansible is invoked upon completion of cluster node provisioning to start tests
  \item Tests are run over ssh inside of Docker on a designated head node and launch mpirun commands
  \item For specific user data processing jobs, the user uploads input data to the chosen cloud storage or cluster NFS server
  \item Ansible or manual ssh performs a fetch of data needed, and then Ansible initiates scientific computation
  \item Upon completion of the application run, the user can fetch data using scp, aws s3 command-line, or other convenient data movement tools
  \item The Terraform destroy command can be used to remove all compute resources and leave data and server image, or all resources entirely
  \item Should a user desire to persist the cluster to restart, cloud provider specific commands can pause/restart the compute resources
\end{enumerate}

\subsection{Deployment Contexts}
\label{sec:deploy}

The public cloud provider we deployed full scientific workflow runs on for this work
is AWS, and we wrote and tested Terraform and Ansible scripts for these 
applications (as well as other scientific workflows) on GCP. 
Our choice of basic network and VM
infrastructure provisioning with Terraform also allows us to support extensions
to other public clouds including Microsoft Azure and many more.
Deployment to other platforms have been explored, but have not been fully
automated at this time.

\subsubsection{Public Cloud}
\label{sec:public}

There are a dizzying array of available cloud provider services, with
accompanying power and convenience in equal measure to the volume of user
documentation and choices presented to a user. Towards the goal of enabling
users to quickly deploy scientific software to run massively parallel or on
a communicating cluster to engage many compute resources, we do not burden
users immediately with undue choices nor present the full complexity of cloud
provider offerings available. Sensible defaults are chosen for the applications
presented and these finer details are exposed in the Terraform
infrastructure provisioning descriptions and can be modified as users see fit,
and as they desire to learn more powerful controls of the performance and
cost of the deployed systems.

The largest differences between the cloud infrastructure providers are the names
of different resources described in the Terraform resource descriptions, written
in a format similar to JSON dubbed HashiCorp Configuration Language (HCL). A GCP
compute network is similar to an AWS virtual private cloud. Both are used to
designate the network the VMs are created on. To control the ingress and egress traffic of the VMs we used 
the AWS security group which is equivalent to the GCP compute firewall. 
In order to facilitate MPI communication for on-demand cluster creation, various
security group port settings in combination with instance level Docker container
settings and internal container process launch commands were attempted to arrive
at working configurations that have been captured in the deployment, build, and
run scripts we present. Cluster access is secured by appropriate default network
controls and ssh access key configuration. To make copies of the base VM in AWS, we 
used the Terraform resource "AMI from instance" to create an image from the base VM and multiple VMs from the
created image. In GCP, we first created a Google compute snapshot of the VM, 
which was then converted to a Google compute disk, which was then used
to create a Google compute image. From there, multiple VMs can be created from
the image. Individual cloud providers tend to host the VM server instances using
industry standard KVM or Xen hypervisors, but any hypervisor derived differences in
execution or performance are beyond the scope of this analysis, with reproducibility
verified at the level of application numerical results.

\subsubsection{Private Cloud}
\label{sec:private}

Aristotle Cloud Federation is a collection of on campus infrastructure hosting resources
at multiple universities that utilize the open source OpenStack infrastructure hosting services
to provide storage, network, and computing services for users. In this work, we deploy
on the OpenStack resources of Cornell University Red Cloud using standard networks
and VMs that are similarly available on other federation sites and OpenStack providers elsewhere.

The largest differences between public and private cloud providers are the Terraform
provider used for infrastructure resource creation, which means different section
names in the desired resource description rendered in HCL. As far as architecture structure 
and resource creation decisions, the network descriptions are again the largest point
of difference, but similarly provide communication to and among the deployed VMs.
Details of private network creation and the many resources necessary to make the first
reachable VM on the infrastructure target are thus automatically created without further
user involvement nor navigation of novel linked webpage deployment instructions.

\section{Practical Outcomes}
\label{sec:ease}

The primary benefit of this work is the ability to quickly and easily move a scientific computing application -- including HPC applications
which are communication-bound -- with its dependencies and associated setup to any number of cloud infrastructures.  This has several
resulting outcomes that are beneficial, with a few disadvantages, and has resulted in many lessons learned.  Our approach increases the 
accessibility of the cloud computing paradigm for scientific computing, has the ability to leverage multi-cloud deployments, adds portability
and reproducibility naturally to the process of scientific software deployment in the cloud, and can simplify the process of iterative software 
development due to rapid deployment options.

Our implementation drastically eases or removes the infrastructure implementation required of scientists and researchers, 
both in understanding and in time to develop, freeing up precious time to target scientific results or performance within an application.
For researchers who do not otherwise have access to large-scale computational resources, or who have access to the cloud but not the
understanding of specific deployment contexts and tools to be able to leverage the cloud effectively, our provided scripts can 
be employed to enable access more readily.  Furthermore, research staff responsible for supporting researchers in scientific computing
can apply this work to on-premise clouds to automate deployments, or to public cloud deployments to expedite researcher progress.

Since Terraform and Ansible are already designed to handle deployments in a variety of clouds, there is no added work to switch from
one cloud to another other than the cloud-specific details that must change, but would need to regardless of deployment method.  The same container
can be used on any of the public cloud vendors and many private clouds.  Furthermore, the same container can be utilized on a personal computer for
development, then deployed in multiple clouds, increasing the ability to push changes and rapidly deploy improvements or new scientific tests.

A disadvantage is if you need to change the implementation, it may require understanding some cloud infrastructure specific details (though it would anyway)
and knowing enough Terraform or Ansible to be able to make the changes.  It is a smaller overall cost than complete manual deployment, but especially a downside 
if you are already familiar with or targeting a single cloud vendor and already have understanding of other tools.  A further disadvantage could be if a container
does not already exist for your application, or one of the cloud services you would like to leverage has not already been scripted, 
then time would need to be spent in development.

The experience of containerizing, automating the deployment of, and running the computations for these scientific applications has
provided a wealth of lessons on both the difficulties and the simplifications available when moving scientific research from a variety
of disciplines to the cloud, which we shall summarize.  The simplest deployment, especially if a Multi-VM setup is required,
can be the best way to illustrate requirements that might otherwise have been overlooked due to the differences between the
cloud computing paradigm and other compute resources.  As with any new system, it pays to start with the simplest use case,
and build up incrementally to the full scale of the application.  For deployment in the cloud, this means not only starting with
a small-scale run, but also a small data input and output, a single VM (as far as possible), a minimal-size container, as so on.
Starting with large data sizes can cause undue complexity to the process of deployment and computation, whether in choosing
the appropriate VM size and type to handle the load, in large long-term storage costs while still developing, or in large egress
charges on public cloud (though this is not an issue for some private campus clouds such as Aristotle Cloud Federation).  Thus,
it is also important to become familiar with the cost model of your chosen cloud provider(s) and determine cost requirements
concurrently with cloud infrastructure requirements.  After a successful run of a small deployment of an application, these
requirements become clearer.

The importance of the choice of software tools one uses cannot be overstated.  While there is a plethora of tools -- whether 
created by specific cloud vendors, industry partners, or otherwise -- that can be used to facilitate configuration, deployment,
automation, and computation in the cloud, it is vital to select the tools that are not only the best for the job, but also enable
the user to get scientific code running quickly.  Across a variety of clouds, we have found that Terraform and Ansible provide 
rapid configuration, deployment, and management of compute resources for scientific workflows in a manner simplified for
those familiar with scripting and similar tools.  Applications using Python, bash scripting, or similar tools are convenient to
run from Ansible, empowering the user to increase automation of the application runtime in the deployed environment with
low effort.

\section{Conclusion}
\label{sec:conclusion}

Initial adoption of cloud computing for deployment of scientific and HPC workflows can require a large lead time to learn new technologies,
develop containers that support software development and production work, comprehension of how to translate requirements to cloud infrastructure options,
and even learning the nuances of how particular cloud vendors operate.  The technologies we've presented in this paper can be very useful tools to
reduce this lead time and deploy scientific runs more rapidly, while increasing reproducibility and portability of the scientific workflows in the process.
We have provided open source code and examples in the hopes that others can leverage this work to increase their own understanding of the cloud
as an infrastructure to support scientific computing, and to deploy new workflows to the benefit of the scientific community at large.

However, moving scientific research applications to the cloud has the potential to be an undertaking, and this is especially true for the aforementioned
HPC applications which are communication-bound.  For the researcher interested in moving an MPI application to the cloud, this
approach may decrease the deployment time and add other benefits, but several other factors are of concern as well in making the 
decision: cost, availability of compute resources, existence of a container template for the application (or time and
knowledge to develop one), and familiarity with the cloud and associated tools.  A compelling case for cloud is when compute cycles are
simply not available elsewhere, or cloud resources (such as campus clouds) are more readily available than HPC-style resources.  Conversely,
the public cloud might be cost-prohibitive for very large-scale or large-output applications, but one is able to secure time on a supercomputer.
More work is needed to fully examine when is a good time to turn to cloud as a resource for scientific computing and HPC applications. 

Our plan for future work includes completing automation of these workflows on Microsoft Azure and Google Cloud, cost analysis of deployments, 
and bare-metal performance comparisons on HPC resources.  There are also more application areas to explore using this approach, including 
those that have need of specialized hardware (such as GPUs).  Furthermore, while we have stated some reasons for selecting this approach in 
favor of others for the deployment of these workflows, a broader analysis of other deployment methods (such as Kubernetes and 
vendor-specific strategies) is another next step.

\begin{acks}

The authors would like to thank the following researchers who contributed to the development of this project: Jim Cordes, Shami Chatterjee, Julianne Quinn, Tristan Shepherd, Robert Wharton, and Marty Sullivan, as well as students supported by the CAC: Elizabeth Holzknecht and Shiva Lakshaman; and Cornell student Shen Wang for FRB\_Pipeline contributions.  The authors would also like to thank the anonymous referees for taking the time to review our work and provide feedback.

This work has been supported by the Cornell University Center for Advanced Computing and the Aristotle Cloud Federation project.
This work is supported by \grantsponsor{NSF}{National Science Foundation}{https://www.nsf.gov} under Grant Number:~\href{https://www.nsf.gov/awardsearch/showAward?AWD_ID=1541215&HistoricalAwards=false}{OAC-1541215}.
Cloud Credits supporting this research were provided via \grantsponsor{AZ}{Amazon AWS Research Credits}{https://aws.amazon.com/research-credits},
\grantsponsor{GCP}{Google Research Cloud Program}{https://edu.google.com/programs/credits/research},
and
\grantsponsor{MS}{Microsoft Azure for Research}{https://microsoft.com/en-us/research/academic-program/microsoft-azure-for-research}.
\end{acks}

\bibliographystyle{ACM-Reference-Format}
\bibliography{pete,astro,lake,wrf,barker,knep}

\end{document}